# The method for solving the KdV-equation

Dmitry Levko

**Abstract**: The method for solving the KdV are considered.

23 May 2007

In [1] the simplest and direct method for construction of the solutions of nonlinear equations was considered. It is allow expressing the solutions of nonlinear equations of special class through the solutions of linear or nonlinear equations with known solutions. Analogous methods already are known for linear and nonlinear Schrödinger equations [2]-[3]. But method are presented in [1] contain some inaccuracies. It don't allow step by step, without some additional desires, to find the solution. In this work the improvement of method [1] are considered.

For demonstration of the modified method the first example from [1] was chosen.

Let us consider the linear equation

$$U_t + U_{xxx} = 0. \tag{1}$$

The new variable in this equation are

$$z = kx + \omega t,$$

where $k$ – wave number, $\omega$ – frequency. Then equation (1) we can rewrite in form:

$$\omega U_z + k^3 U_{zzz} = 0. \tag{2}$$

The solution of such equation are considered in form

$$U(z) = A \cdot \sin(z), \tag{3}$$

where $A$ some constant. The dispersion correlation of (2) is next

$$\omega = k^3. \tag{4}$$

From classical mechanics are known that the ordinary differential equation of motion can be represent in form

$$U_{zz} = -\frac{\omega}{k^3}U + \beta = \frac{\partial V(U)}{\partial U}, \tag{5}$$

where $V(U)$ – is the potential in which the image point is move. From (3) and (2) we can find that $\beta = 0$. Then we define the potential after the one integration. It is:

$$V(U) = -\frac{\omega}{2k^3}U^2 + \lambda. \tag{6}$$

Here $\lambda$ – some constant.

Now let us consider the KdV

$$Q_t + 6QQ_x + Q_{xxx} = 0. \tag{7}$$

The new variable

$$z = kx - \omega' t.$$

Analogous operations as for (1) can be made for (7). With new variable (7) are rewritten

$$-\omega' Q_z + 6kQQ_z + k^3 Q_{zzz} = 0. \tag{8}$$

The potential of this equation has the form

$$\overline{V}(Q) = \frac{\omega'}{2k^3}Q^2 - \frac{1}{k^2}Q^3 + \overline{\beta}Q + \overline{\lambda}, \tag{9}$$

where $\overline{\beta}$ and $\overline{\lambda}$ – constants. From asymptotical equality to zero on infinity the function $Q(z)$ and it's derivatives are got $\overline{\beta} = \overline{\lambda} = 0$.

To express the solution of (7) through the solution (1) let us assume that both solutions are related by the correlation:

$$U = g(Q). \tag{10}$$

The function $g(Q)$ can be finding from next expression [4]

$$g'(Q) = \sqrt{\frac{V(g)}{\overline{V}(Q)}}. \tag{11}$$

Then the system (1), (7) is given the equality

$$\int \frac{dg}{\sqrt{\lambda - \frac{\omega}{2k^3}g^2}} = \int \frac{dQ}{\sqrt{\frac{\omega'}{2k^3}Q^2 - \frac{1}{k^2}Q^3}}. \quad (12)$$

The solution of (12) in connect with (3) is determined the functions $g$ and $Q$.

The first integral

$$\sqrt{\frac{2k^3}{\omega}} \arcsin \frac{g}{\sqrt{\frac{2k^3 \lambda}{\omega}}},$$

and the second

$$\sqrt{\frac{2k^3}{\omega'}} \ln \left| \frac{1 - \sqrt{1 - \frac{2kQ}{\omega'}}}{1 + \sqrt{1 - \frac{2kQ}{\omega'}}} \right|.$$

We can find the constant $\lambda$ if we equate (3) to expression for function $g$ that we got from (12)

$$\lambda = \frac{A^2}{2}.$$

The equality of phases for $\sin$ is allowed to determine the solution of (7)

$$Q(z) = \frac{\omega'}{2k} \cdot \operatorname{sech}^2 \left( \frac{z}{2} \sqrt{\frac{\omega'}{\omega}} \right).$$

To get the known soliton solution we need in last expression to suppose next dispersion correlation

$$\omega' = 4k^3. \quad (13)$$

Then finally

$$Q(z) = 2k^2 \cdot \operatorname{sech}^2 k(x - 4k^2 t). \quad (14)$$

The given argumentations can be apply to all examples in [1]. It'll give the same results.

### Literature
1. D. Bazeia, A. Das, L. Losano and A. Silva. Arxiv: nlin.SI / 0703035v1
2. G. L Lamb, Jr. Elements of Solitons Theory. Mir, Moscow, 1983 (in Russian)
3. D. Levko. arXiv: nlin.SI / 0612027
4. D. Bazeia, L. Losano, and J.M.C. Malbouisson, Phys. Rev. D **66**, 101701(R) (2002)

Dmitry Levko: Institute of Physics National Ukrainian Academy of Sciences
d.levko@gmail.com